# Demonstration of Rashba spin splitting in GaN-based heterostructures

W. Weber[1], S.D. Ganichev[1], Z.D. Kvon[2], V.V. Bel'kov[3], L.E. Golub[3], S.N. Danilov[1], D. Weiss[1], W. Prettl[1], Hyun-Ick Cho[4] and Jung-Hee Lee[4]

[1] Fakultät Physik, University of Regensburg, 93040, Regensburg, Germany
[2] Institute of Semiconductor Physics, Novosibirsk, 630090, Russsia
[3] A.F. Ioffe Physico-Technical Institute, Russian Academy of Sciences, 194021 St. Petersburg, Russia
[4] Kyungpook National University, 1370, Sankyuk-Dong, Daegu 702-701, Korea

Abstract: The circular photogalvanic effect (CPGE), induced by infrared radiation, has been observed in (0001)-oriented $n-$GaN quantum well (QW) structures. The photocurrent changes sign upon reversing the radiation helicity demonstrating the existence of spin-splitting of the conduction band in $k$-space in this type of materials. The observation suggests the presence of a sizeable Rashba type of spin-splitting, caused by the built-in asymmetry at the AlGaN/GaN interface.

Gallium nitride is a potentially interesting material system for spintronics since it is expected to become ferromagnetic with a Curie-temperature above room temperature if doped with manganese [1]. Long spin relaxation times observed in this materials are another promising property for possible applications [2]. Little is known so far about spin orbit interaction in GaN based heterojunctions like existence of Rashba spin-splitting in the band structure which would provide a potential handle for spin manipulation [3]. Strong spin-orbit effects are expected to be in narrow-gap materials only [4]. A large piezoelectric effect which causes a strong electric field at the $Al_xGa_{1-x}N$/GaN interface and the strong polarization induced doping effect, on the other hand, may result in a sizeable Rashba contribution to spin-splitting of the band due to spin-orbit interaction (~ 1 meV) [5]. Indeed a spin-splitting of 9 meV was extracted from beatings of Shubnikov-de-Haas oscillations in $Al_{0.25}Ga_{0.75}N$/GaN heterostructures [6]. However, such beatings were ascribed to magneto-intersubband scattering by others [7]. On the other hand the observation of short spin relaxation times was attributed to D'yakonov Perel mechanism which requires Rashba spin splitting [8]. To investigate the presence of a sizeable spin-splitting in this material class we investigate the circular photogalvanic effect (CPGE) [9,10,11].

The CPGE as well as $k$-linear spin splitting of the band structure are only permitted in gyrotropic media. In such materials a linear relation between polar vectors (electric current $j$, quasi-momentum $q$ etc.) and axial vectors (photon angular momentum, spin etc.) is allowed by symmetry. Both GaN/AlGaN QW structures and bulk GaN belong to the family of wurtzite-type semiconductors which are gyrotropic. As were pointed out in [12,13] in these media the spin-orbit part of the Hamiltonian has the form

$$\hat{H}_{SO} = \alpha[\boldsymbol{\sigma} \times \boldsymbol{k}]_z. \tag{1}$$

Here the $z$-axis is directed along the hexagonal $c$-axis and $\boldsymbol{\sigma}$ is the vector of Pauli matrices. In bulk wurtzite materials the constant α in the Hamiltonian (1) is solely due to bulk inversion asymmetry (BIA). In heterostructures, an additional source of $k$-linear spin splitting, induced by structure inversion asymmetry (SIA), exists. It occurs in asymmetric semiconductor heterostructures of any material [3]. If both, bulk and structure asymmetries, are present the resulting coupling constant $\alpha$ is equal to the sum of BIA and SIA contributions to the spin-orbit part of the Hamiltonian. Note that in bulk III-V semiconductors which lack the hexagonal $c$-axis the constant $\alpha$ is zero.

To probe the presence of a sizable spin-splitting below we employ the CPGE, which is sensitive to spin splitting of the band structure. By optical excitation with circularly polarized light spin-up and spin-down subbands get non-uniformly populated in $k$-space. This is a consequence of optical selection rules and energy and momentum conservation, leading to a current which is probed in experiment [9,10,11]. The fingerprint of the CPGE photocurrent is its dependence on the helicity of the radiation field. The current reverses its direction by switching the light's polarization from right-handed circular to left-handed circular and vice versa.

The experiments were carried out on $Al_{0.3}Ga_{0.7}N/GaN$ heterojunctions grown by MOCVD on a C(0001)-plane sapphire substrate. The thickness of the AlGaN layers was varied between 30 nm and 100 nm. An undoped 33 nm thick GaN buffer layer, which was grown under a pressure of 40 Pa at 550 °C, is followed by an undoped GaN layer (2.5 μm), grown under a pressure of 40 Pa at 1025 °C. The undoped $Al_{0.3}Ga_{0.7}N$ barrier was grown at 6.7 Pa and a temperature of 1035 °C. The electron mobility in the 2DEG was typically about 1200 cm$^2$/Vs at electron density (1-1.4)·10$^{13}$ cm$^{-2}$ at room temperature. To measure the photocurrent two pairs of ohmic contacts have been centred along opposite sample edges. The experiments were carried out in two different spectral ranges: the mid-infrared (MIR) regime with wavelength between 9.2 μm and 10.8 μm and the terahertz (THz) regime at 77 μm, 90.5 μm, 148 μm, 280 μm and 496 μm. The latter wavelengths were achieved by an optically pumped molecular $NH_3$ laser [14] while a pulsed TEA $CO_2$ laser and a commercial Q-switched $CO_2$ laser with a peak power of about 1 kW were employed in the MIR-regime. While the THz radiation causes indirect Drude-like optical transitions in the lowest subband of the 2DEG, the MIR radiation can additionally induce direct optical transitions between the subbands. To obtain circularly polarized radiation needed for CPGE the laser light was passed through a Fresnel rhombus or quartz λ/4 plates for MIR and THz radiation, respectively. The helicity $P_{circ} = \sin 2\varphi$ of the incident light was varied from −1 (left handed circular, σ$_-$) to +1 (right handed circular, σ$_+$). Here $\varphi$ is the angle between the initial polarization plane and the optical axis of the λ/4 plate or the Fresnel rhombus. The current $j$ generated by the circularly polarized light in the unbiased samples was measured at room and liquid nitrogen temperatures via the voltage drop across a 50 Ω load resistor in a closed circuit configuration. The voltage was recorded with a storage oscilloscope.

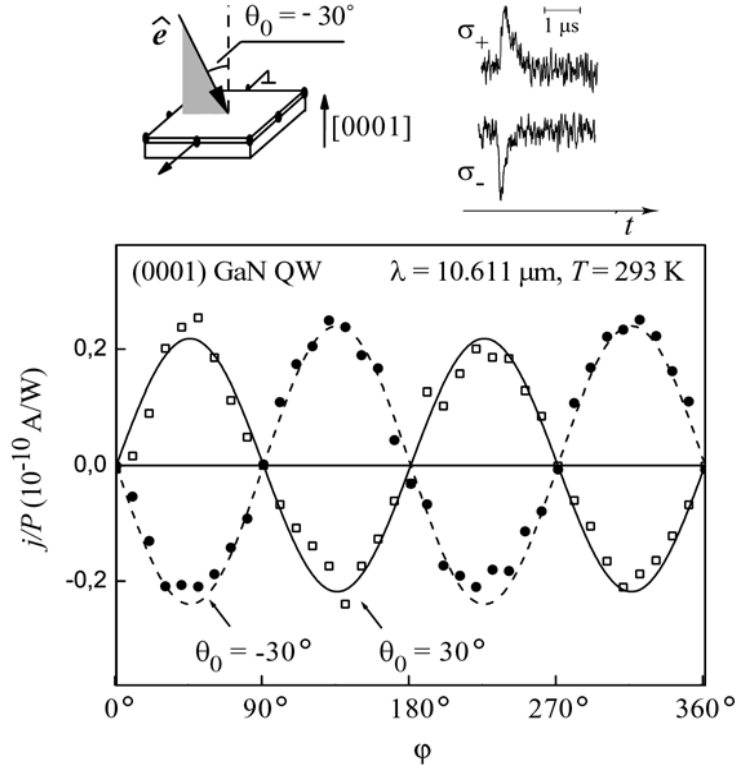

*Fig. 1. Photocurrent in GaN QWs normalized by the radiation power P as a function of the phase angle $\varphi$ defining helicity. Measurements are presented for room temperature and irradiation by light of Q-switched $CO_2$ laser at the wavelength $\lambda = 10.611$ µm. The current $j_x$ is measured for direction perpendicular to propagation of light (angle of incidence $\Theta_0 = \pm 30°$). Solid and dashed lines show ordinate scale fits according to Eqns. (2)-(5). Insets show the geometry of the experiment and the temporal structure of the current response for right- and left-circularly polarized radiation.*

Irradiating the (0001) AlGaN/GaN heterojunction by circularly polarized light at oblique incidence, as sketched in the inset of Fig. 1, causes a photocurrent signal measured across a contact pair. The measured current follows the temporal structure of the applied 40 ns laser pulses and is shown in the insets in Fig. 1 for two polarization states. The current reverses its direction by switching the sign of the radiation helicity (see Fig. 1). The fact that the current is proportional to the radiation helicity proves the circular photogalvanic effect as origin of the photocurrent. The signal proportional to the helicity is only observed under oblique incidence. This is shown in Fig. 2 where, for fixed helicity, the angle of incidence, $\Theta_0$, is varied. The current vanishes for normal incidence and changes its polarity. The photocurrent in the layer flows always perpendicularly to the direction of the incident light propagation and its magnitude does not change by rotating the sample around the growth axis. All characteristic features persist from 77 to 300 K. The observed photocurrents have the same order of magnitude as those measured in GaAs, InAs, and SiGe QWs [10].

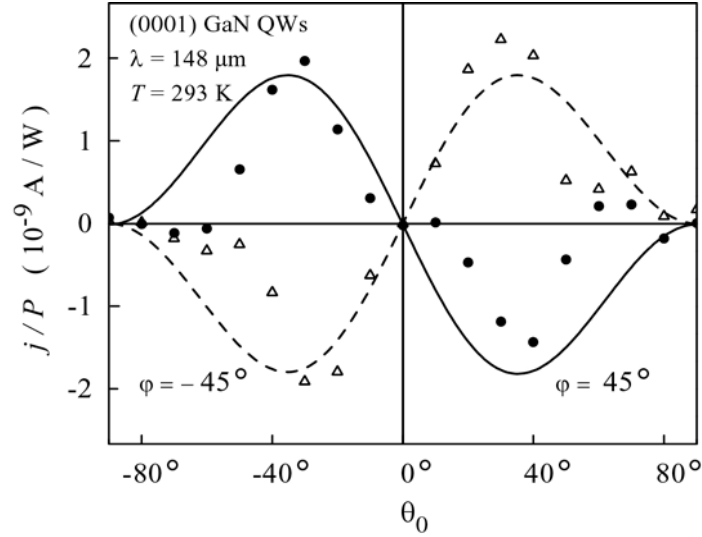

*Fig. 2. Photocurrent in GaN QWs normalized by the light power P as a function of the angle of incidence $\Theta_0$ for right- and left-handed circularly polarized terahertz radiation measured perpendicular to light propagation (T = 296 K, λ = 148 μm). Solid and dashed lines show ordinate scale fits according to Eqns. (2)-(5).*

The effect is observed for all wavelengths used between 9 μm and 496 μm. Data in Fig. -1 shows the effect for a wavelength of 10.6 μm using a Q-switched $CO_2$ laser. Data for λ=148 μm are shown in Fig. 2. While the overall signature is the same the strength of the photocurrent depends on the wavelength (see Fig. 3). The spectral dependence of the CPGE in the THz range (λ ≥ 77 μm) agrees with the expected behaviour of the CPGE for indirect (Drude-like) transitions. However the rapid resonance-like increase of the signal at short wavelengths (see inset of Fig. 3) obtained with the $CO_2$ laser (9.2-10.8 μm) cannot be explained by this mechanism. We ascribe this spectral dependence to resonant direct intersubband optical transitions between the ground and the first excited size-quantized subbands. In GaN quantum wells which are not yet sufficiently investigated, spectroscopic data are still not available. However, we estimate that the energy separation between the two lowest subbands should be about 150 meV in the approximation of the triangular quantum well. This value slightly exceeds the photon energies used in our experiments. It is not surprising because the real shape of the quantum well is nonlinear and strength of the electrical field significantly falls as the distance from a surface increases. More accurate comparison needs self-consistent solution of Schrödinger and Poisson's equations. In addition to the CPGE current detected in the direction normal to the in-plane wave vector of radiation a signal is also observed along the in-plane propagation direction. This signal has equal magnitude and the same sign for right- and left-handed circularly polarized radiation and is ascribed to the linear photogalvanic effect and the photon drag effect [10].

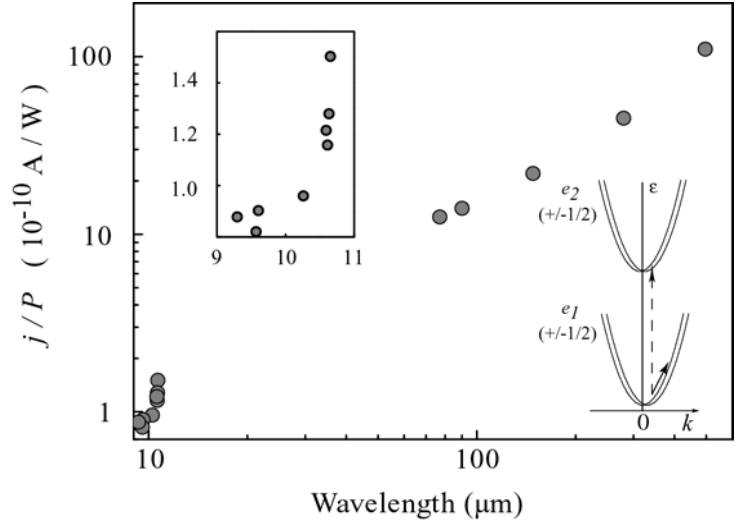

*Fig. 3. Spectral dependence of the CPGE current in GaN QWs at room temperature. Sketch demonstrates possible optical transitions.*

The observed dependence of the photocurrent on helicity and angle of incidence can be described by phenomenological theory [10,11] adapted to wurtzite-type low-dimensional systems. Spin-orbit interaction Eq. (1) leads to a CPGE current in the plane of a AlGaN/GaN heterostructure given by

$$\bm{j} = P_{circ}\, \gamma\, \hat{\bm{e}}_\|. \qquad (2)$$

Here $\bm{j}$ is the net current density, $P_{circ}$ and $\hat{\bm{e}}_\|$ are the degree of circular polarization and the projection onto the QW plane of the unit- vector $\hat{\bm{e}}$ pointing in the direction of light propagation, respectively. The second-rank pseudotensor $\gamma$ is proportional to the spin-orbit constant $\alpha$. It has two nonzero components and is described by one linearly-independent constant:

$$\gamma_{xy} = -\gamma_{yx}, \quad \gamma_{ii} = 0. \qquad (3)$$

The dependence of the CPGE on the angle of incidence $\Theta_0$ is determined by the value of the projection $\hat{e}_\|$ which is for the excitation say in the plane of incidence $yz$ is given by

$$\hat{e}_P = \hat{e}_y = t_p t_s \sin\Theta, \qquad (4)$$

where $\Theta$ is the refraction angle defined by $\sin\Theta = \sin\Theta_0 / n_\omega$, and $t_p$ and $t_s$ are transmission coefficients for linear $p$ and $s$ polarizations, respectively. They are given by Fresnel's formula:

$$t_p t_s = \frac{4\cos^2\Theta_0}{(\cos\Theta_0 + \sqrt{n_\omega^2 - \sin^2\Theta_0})(n_\omega^2\cos\Theta_0 + \sqrt{n_\omega^2 - \sin^2\Theta_0})}. \qquad (5)$$

Equations (2) - (5) fully describe the experimental observations. Corresponding calculations are shown as solid lines in Figs. 1 and 2. The current follows the radiation helicity and reverses its sign upon reversing the direction of incidence, $\hat{e}_P \to -\hat{e}_P$. In contrast to zinc-blende based systems like GaAs QWs the CPGE in wurtzite based systems (with the hexagonal *c*-axis normal to the QW) should be independent of the in-plane propagation direction $\hat{e}_\|$. This prediction of the phenomenological theory is observed in experiment as discussed above.

The observation of the CPGE with strength comparable to that observed in GaAs and InAs QWs unambiguously demonstrates a substantial Rashba splitting of spin subbands in GaN heterojunctions. In contrast to zinc-blende based III-V QWs, where interference of BIA and SIA results in varying angles between electron spin and its momentum, *k*, for different crystallographic directions, the electron spin in GaN QWs is always perpendicular to *k*. This is demonstrated by our experiments where the CPGE current always flows perpendicular to $\hat{e}_P$ and does not change its amplitude if the in-plane direction is varied. The reason of this axial isotropy is that both, BIA and SIA, lead to the same form of spin-orbit interaction given by Eq. (1).

Acknowledgements: This work is supported by the DFG, RFBR, INTAS, RAS „Dynasty" Foundation — ICFPM and by the grant N R01-2003-000-10769 (2004), BK 21 of Korea Science and Engineering Foundation.

References

[1] T. Dietl, H. Ohno, F. Matsukura, J. Cibert, and D. Ferrand, Science **287**, 1019 (2000).

[2] B. Beschoten, E. Johnston-Halperin, D. K. Young, M. Poggio, J. E. Grimaldi, S. Keller, S. P. DenBaars, U. K. Mishra, E. L. Hu, and D. D. Awschalom, Phys. Rev. B **63**, 121202 (2001).

[3] *Semiconductor Spintronics and Quantum Computation*, eds. D.D. Awschalom, D. Loss, and N. Samarth, in the series *Nanoscience and technology*, eds. K. von Klitzing, H. Sakaki, and R. Wiesendanger (Springer, Berlin, 2002).

[4] R. Winkler, *Spin-orbit coupling effects in two-dimensional electron and hole systems*, in *Springer Tracts in Modern Physics*, Vol.191 ( Springer-Verlag, Berlin, 2003).

[5] V. I. Litvinov, Phys. Rev. B **68**, 155314 (2003).

[6] Ikai Lo, J. K. Tsai, W. J. Yao, P. C. Ho, Li-Wei Tu, T. C. Chang, S. Elhamri, and W. C. Mitchel, K. Y. Hsieh, J. H. Huang, H. L. Huang, and Wen-Chung Tsai, Phys. Rev. B **65**, 161306 (2002).

[7] N. Tang, B. Shen, Z.W. Zheng, J. Liu, D.J. Chen, J. Lu, R. Zhang, Y. Shi, Y. D. Zheng, Y.S. Gui, C.P. Jiang, Z.J. Qiu, S.L. Guo, J.H. Chu, K. Hoshino, T. Someya, and Y. Arakawa J. Appl. Phys. **94**, 5420 (2003)

[8] I. A. Buyanova, J.P. Bergman, W.M. Chen, G. Thaler, R. Frazier, C.R. Abernathy, S.J. Pearton, Jihyun Kim, F. Ren, F.V. Kyrychenko, C.J. Stanton, C.-C. Pan, G.-T. Chen, J.-I. Chyi, and J.M. Zavada, J. Vac. Sci. Technol. B **22**, 2668 (2004).

[9] S.D. Ganichev, E. L. Ivchenko, S.N. Danilov, J. Eroms, W. Wegscheider, D. Weiss, and W. Prettl, Phys. Rev. Lett. **86**, 4358 (2001).

[10] S.D. Ganichev and W. Prettl, J. Phys.: Condens. Matter **15**, R935 (2003).

[11] E.L. Ivchenko, *Optical Spectroscopy of Semiconductor Nanostructures*, (Alpha Science Int., Harrow, UK 2005).

[12] É. I. Rashba, Fiz. Tverd. Tela **2**, 1224 (1960) [Sov. Phys. Solid State **2**, 1109 (1960)].

[13] Yu.A. Bychkov and E.I. Rashba, Pis'ma Zh. Eksp. Teor. Fiz. **39**, 66 (1984) [JETP Lett., **39**, 78 (1984)].

[14] S.D. Ganichev, I.N. Yassievich, and W. Prettl, J. Phys.: Condens. Matter **14**, R1263 (2002).